\documentclass[12pt]{article}

\usepackage{latexsym}
\usepackage[bf,footnotesize]{caption}   
\begin{document}
\setlength{\unitlength}{1mm}

\newcommand{\be}{\begin{equation}}
\newcommand{\ee}{\end{equation}}
\newcommand{\ba}{\begin{eqnarray}}
\newcommand{\ea}{\end{eqnarray}}
\newcommand{\ban}{\begin{eqnarray*}}
\newcommand{\ean}{\end{eqnarray*}}

\newcommand{\e}{{\mathrm e}}
\newcommand{\bbox}{\bar{\phantom{!}\Box\phantom{!}}}
\newcommand{\Tr}{\phantom{!}\mbox{\bf Tr}\phantom{!}}
\newcommand{\tr}{\phantom{!}\mbox{\bf tr}\phantom{!}}
\newcommand{\ra}{\rangle}
\newcommand{\la}{\langle}

\newcommand{\n}[1]{\label{#1}}
\newcommand{\eq}[1]{Eq.(\ref{#1})}
\newcommand{\ind}[1]{\mbox{\bf\tiny{#1}}}
\renewcommand\theequation{\thesection.\arabic{equation}}

\newcommand{\nn}{\nonumber \\ \nonumber \\}
\newcommand{\nl}{\\  \nonumber \\}
\newcommand{\pr}{\partial}
\renewcommand{\vec}[1]{\mbox{\boldmath$#1$}}

\title{{\hfill {\small Alberta-Thy-20-02 } } \vspace*{2cm} \\
Quantum Radiation from a $5$-Dimensional Rotating Black Hole}
\author{\\
Valeri Frolov${}^1$ and Dejan Stojkovi\'{c}${}^1$}
\maketitle
\noindent  {
$^{1}${ \em
Theoretical Physics Institute, Department of Physics, \ University of
Alberta, \\ Edmonton, Canada T6G 2J1}
\\
e-mails: frolov@phys.ualberta.ca, dstojkov@phys.ualberta.ca
}
\bigskip

\date{\today}
\maketitle
\noindent

\begin{abstract}
We study a massless scalar field propagating in the background
of  a five-dimensional rotating black hole. We showed that in
the Myers-Perry metric describing such a black hole the massless
field equation  allows the separation of variables. The obtained
angular equation is a generalization of the equation for
spheroidal functions. The radial equation is similar to the
radial Teukolsky equation for the 4-dimensional Kerr metric. We
use these results to quantize the massless scalar field in the
space-time of the 5-dimensional rotating black hole and to derive
expressions for energy and angular momentum fluxes from such a
black hole.
\end{abstract}

\vspace{.3cm}


\section{Introduction}\label{s1}
\setcounter{equation}0


Solutions of Einstein's equations describing black holes in
space-times with more than usual $(3+1)$-dimensions have been
known in literature for forty years \cite{Tan}. Until recently,
they were practically only of academic interest. The situation
has changed with appearance of so-called brane world models
\cite{BW} in which large extra spatial dimensions can exist
(much larger than the apparent Planck length scale $\sim
10^{-33}$cm). In this framework, the fundamental quantum gravity
energy scale could be in $1$TeV range, while the characteristic
length scale (compactification radius of extra dimensions) can be
as large as $0.1$mm. This allows existence of higher dimensional
mini-black holes which can be described within the classical
theory of gravity. Being gravitational solitons black holes in
the brane world can `live' out of the brane in the bulk space. If
the gravitational radius of such a black hole is much smaller
than the distance to the brane and the characteristic length
defined by the bulk curvature and/or the size of extra dimensions,
the influence of the external conditions on the properties of the
space-time near the horizon is small. Under these conditions in
order to describe the black hole one can use solutions of vacuum
Einstein equations. This also can be true for small black holes
attached to the brane provided the brane is ``soft", that is its
tension is small.

From phenomenological point of view, the most exciting
possibility is that such mini-black holes can be produced in
particle collisions in near future accelerator  and cosmic ray
experiments \cite{acc}. In order to predict the correct
experimental signature of these events, one has to know the basic
properties of solutions of these  higher dimensional black holes.
In general case, the impact parameter in particle collision will
be non-zero. Therefore, majority of black holes produced in such
way would be rotating.

Higher dimensional black holes were studied before in different
contexts. For example, supersymmetric rotating black holes were studied in \cite{GibHer}
and  \cite{Her}. The solution analyzed there is not a vacuum solution
of Einstein's equations and requires some special choice of
parameters in order accommodate the supersymmetry. In \cite{Cvetic1,Cvetic2},
rotating black holes were studied in the context of string theory.
Scalar field gray body factors were calculated for a very general
case of a boosted vacuum solution of Einstein's equations.

The metric for higher dimensional rotating black holes was first
derived by Myers and Perry \cite{MyPe:86}. In the present paper, we
study  the massless scalar field equation in the background of a five-dimensional
rotating black hole described by the Myers-Perry metric. We show
that this equation allows the  separation of variables. The
obtained angular equation is a generalization of the equation for
spheroidal functions. The radial equation is similar to the
radial Teukolsky equation for the 4-dimensional Kerr metric.

In the case of $3+1$ dimensions, there is only one parameter of
rotation and there is an axis of rotation which stays invariant
under rotation.  The rotation group in $4+1$ dimensions, $SO(4)$,
has two Casimir operators. Rather than an axis of rotation, there
exist  planes of rotation which stay invariant under the
rotation. This implies that, in general, there are two parameters
of rotation corresponding to two independent planes of rotation.
In a special case, one can set one of the parameters to zero and
consider rotations only in one plane. We find interesting that in
another special case, when both parameters of rotation are
non-zero and equal in magnitude, the angular equation reduces to
a case of a non-rotating black hole. For this special case the
space-time has two additional Killing vectors. It is interesting
that the spatial 3-dimensional slices of this space-time are
homogeneous and belong to the Bianchi type VIII class.

We analyze the structure and asymptotics of solutions of the
radial equation which determine the black hole gray-body
factors.  We use these results to quantize the massless scalar
field in the space-time of the 5-dimensional rotating black hole
and to derive expressions for energy and angular momentum fluxes
from such a black hole.

\section{Myers-Perry Metric}\label{sec:s2}
\setcounter{equation}0

\subsection{Generic case}

Following Myers and Perry \cite{MyPe:86} we write the metric of a
$5$-dimensional rotating black hole in the form
\[
ds^2=-dt^2+(r^2+a_1^2)(d\mu_1^2+\mu_1^2\, d\phi_1^2)+
(r^2+a_2^2)(d\mu_2^2+\mu_2^2\, d\phi_2^2)
\]
\be
\n{1.1}
+{\Pi\, {\cal F}\over \Pi-r_0^2\, r^2}\, dr^2+ {r_0^2\,
r^2\over \Pi\, {\cal F}}\, (dt+a_1\, \mu_1^2\, d\phi_1+ a_2\, \mu_2^2\,
d\phi_2)^2\, ,
\ee
\be\n{1.2}
{\cal F}=1-{a_1^2\, \mu_1^2\over
r^2+a_1^2}-{a_2^2\, \mu_2^2\over r^2+a_2^2}\, ,
\ee
\be\n{1.3}
\Pi=(r^2+a_1^2)(r^2+a_2^2)\, .
\ee
Here $r_0$ is length parameter
connected with the black hole mass $M$ \be\n{1.4} M= \frac{3
r_0^2}{8 \sqrt{\pi}G}  \ee where $G$ is the
$(4+1)$-dimensional gravitational coupling constant.
Besides $r_0$ the metric (\ref{1.1}) contains two rotation parameters,
$a_1$ and $a_2$. The variables $\mu_1$ and $\mu_2$ are not
independent. They obey a constraint \be\n{1.5}
\mu_1^2+\mu_2^2=1\, . \ee

Instead of keeping the symmetric form of the metric (\ref{1.1}) we
prefer to solve the constraint (\ref{1.5})  explicitly. We use the
following parametrization
\be\n{1.6}
\mu_1=\sin\theta\, ,\hspace{1cm}\mu_2=\cos\theta\, .
\ee
Let us also introduce the following notations
\be\n{1.7}
a=a_1\, ,\hspace{0.5cm} b=a_2\, ,\hspace{0.5cm}
\phi=\phi_1\, ,\hspace{0.5cm}\psi=\phi_2\, ,
\ee
\be\n{1.8}
\rho^2=r^2+a^2\,\cos^2\theta+b^2\,\sin^2\theta\, ,
\ee
\be\n{1.9}
\Delta=(r^2+a^2)(r^2+b^2)-r_0^2\, r^2\, .
\ee
Then the metric (\ref{1.1}) takes the form
\be\n{1.10}
ds^2=d\gamma^2+{r^2\rho^2\over \Delta}\, dr^2+\rho^2\, d\theta^2\, .
\ee
\[
d\gamma^2\equiv \gamma_{AB}\, dx^A\, dx^B=- dt^2 +(r^2+a^2)\, \sin^2\theta\, d\phi^2+
(r^2+b^2)\, \cos^2\theta\, d\psi^2
\]
\be\n{1.11} + {r_0^2\over \rho^2} \left[dt+a\, \sin^2\theta\,
d\phi +b\, \cos^2\theta\, d\psi  \right]^2\, . \ee Here
$A,B=0,3,4$ and $x^0=t$, $x^3=\phi$, $x^4=\psi$. Angles $\phi$ and
$\psi$ take values from the interval $\left[0,2\pi \right]$,
while angle $\theta$ takes values $\left[0,\pi/2 \right]$.

The metric (\ref{1.10}) is invariant under the following
transformation
\be\n{2.0a}
a \leftrightarrow b\, ,\hspace{0.5cm}
\theta \leftrightarrow {\pi\over 2}-\theta\, ,\hspace{0.5cm}
\phi \leftrightarrow \psi \, .
\ee
It possesses 3 Killing vectors, $\partial_t$,
$\partial_{\phi}$ and $\partial_{\psi}$. For this metric
\be\n{2.1a}
\sqrt{-g}=\sin\theta\cos\theta\, r\, \rho^2\, .
\ee
The black hole horizon is located at $r=r_+$ where
\be\n{2.2a}
r_{\pm}^2={1\over 2}\left[r_0^2-a^2-b^2\pm
\sqrt{(r_0^2-a^2-b^2)^2-4a^2b^2}\right]\, .
\ee The angular
velocities $\Omega_a$ and $\Omega_b$ and the surface gravity
$\kappa$ are
\be\n{2.2b}
\Omega_a={a\over r_+^2+a^2}\, ,
\hspace{1cm} \Omega_b={b\over r_+^2+b^2}\, \ee \be\n{2.2bb}
\kappa= \left. {\partial_r\Pi -2r_0^2\, r \over 2r_0^2\,
r^2}\right|_{r=r_+} \, .
\ee

\subsection{Degenerate case} \label{DC}

As we already mentioned in a general case the Myers-Perry metric
has 3 Killing vectors. The space-time becomes more symmetric when
$a=b$. To demonstrate this let us consider first the geometry of
the section $t=$const. It has the form
\be
ds_4^2={r^2\,
(r^2+a^2)\over (r^2+a^2)^2-r_0^2\, r^2}\, dr^2+ds_3^2\, ,
\ee
\be\n{FF.1}
ds_3^2= \alpha\, \left( d\theta^2+ \sin^2\theta\,
d\phi^2+\cos^2\theta\, d\psi^2 \right) +\beta\left(
\sin^2\theta\, d\phi +\cos^2\theta\, d\psi\right)^2\, ,
\ee
where
\be
\alpha=r^2+a^2\, ,\hspace{1cm} \beta={r_0^2\, a^2\over
r^2+a^2}\, .
\ee
By introducing new coordinates \be
\Phi=\psi-\phi\, ,\hspace{1cm} \Psi=\psi+\phi\, ,\hspace{1cm}
\vartheta= 2\theta\, , \ee where $\Psi$  takes values from the
interval $\left[0,4\pi \right]$, $\Phi$ from $\left[-2\pi,2\pi
\right]$ and $\vartheta$ from $\left[0,\pi \right]$, one can rewrite
the metric (\ref{FF.1}) as follows \be ds_3^2= A\, (d\vartheta^2+
\sin^2\vartheta\, d\Phi^2)+B\, (\cos\vartheta\, d\Phi+d\Psi)^2\, . \ee
Here,
 \be A={\alpha\over 4}\, ,\hspace{1cm} B={\alpha+\beta\over 4}\, . \ee
This is a
canonical form of the Bianchi VIII type metric when there exists a
four-parameter group  of isometries acting transitively  in the
3-dimensional space \cite{Petrov}\footnote{In \cite{Galtsov}
rotating black hole solutions to four-dimensional
Einstein-Maxwell-dilaton-axion gravity were analyzed. It was
shown that they are closely
related to the 5-dimensional rotating Myers-Perry black hole with $a =
b$.}. The corresponding Killing vectors are
\[
K_1 = \partial_{\Phi}\, ,\hspace{1cm} K_2=\partial_{\Psi}\, ,
\]
\be K_3=\cos\Phi\, \partial_{\vartheta}-\cot\vartheta\, \sin\Phi\,
\partial_{\Phi}+{\sin \Phi\over \sin\vartheta}\, \partial_{\Psi}\, ,
\ee
\[
K_4=-\sin\Phi\, \partial_{\vartheta}-\cot\vartheta\, \cos\Phi\,
\partial_{\Phi}+{\cos\Phi\over \sin\vartheta}\, \partial_{\Psi}\, .
\]
It is easy to check that these vectors (together with
$\partial_t$) are also Killing vectors for the 5-dimensional
space-time metric (\ref{1.10}) when $a=b$.

Killing vectors can be viewed as generators of the symmetry group
of the manifold. We can obtain a more familiar form of these
Killing vectors. First note that:

\be K_3 K_4 -K_4 K_3 = K_1 \ee where multiplications denote
successive applications of an operator.
 Then, the quadratic combination $K_3^2+K_4^2 +K_1^2$ yields

 \be
K_3^2+K_4^2 +K_1^2 = \frac{\partial^2}{\partial \vartheta^2}
+\cot \vartheta \frac{\partial}{\partial \vartheta}
+\frac{1}{\sin^2 \vartheta }\left( \frac{\partial^2}{\partial \Phi^2}
-2 \cos \vartheta \frac{\partial^2}{\partial \Phi \partial \Psi}
+\frac{\partial^2}{\partial \Psi^2} \right)
 \ee
We can identify
 \be\n{ang} J_{\Phi} \equiv -i K_1 \ , \ \ \ \ \ J_{\Psi}
\equiv -i K_2 \ , \ \ \ \ \ J^2 \equiv -(K_3^2+K_4^2+K_1^2) \ ,
\ee where $J_{\Phi} , J_{\Psi}$ and $J^2$ are familiar angular
momentum operators, while $\Phi, \Psi$ and $\vartheta$ are Euler
angles for the rotation group $O(3)$.

The scalar curvature $R$ for the metric (\ref{FF.1})
\be
R = 2\frac{3(r^2+a^2)^2-a^2 r_0^2}{(r^2+a^2)^3}
\ee
is constant at fixed $r$ and has a positive
sign as long as $r > a \sqrt{\frac{\sqrt{3}}{3} \frac{r_0}{a}
-1}$.

Another interesting observation is the following. Instead of
metric on slices $t=$const, one can consider the 4-dimensional
foliation of the space-time by the Killing trajectories of the
field $\xi_t=\partial_t$. The metric on this foliations is
determined as \cite{Geroch} \be
g^{(4)}_{\mu\nu}=g_{\mu\nu}-{\xi_{\mu}\xi_{\nu}\over \xi^2}\, .
\ee For the Myers-Perry metric (\ref{1.1}) this metric is \be
dS^2\equiv g^{(4)}_{\mu\nu}dx^{\mu}\, dx^{\nu}= {r^2\,
(r^2+a^2)\over (r^2+a^2)^2-r_0^2\, r^2}\, dr^2 +dS_3^2\, , \ee
where $dS_3^2$ has the form (\ref{FF.1}) with \be
\alpha=r^2+a^2\, ,\hspace{1cm} \beta={r_0^2\, a^2\over
r^2+a^2-r_0^2}\, . \ee Thus this metric is also a metric of
homogeneous 3-dimensional space with the 4-parameter group of motion.

The scalar curvature for the metric $dS_3^2$
\be R = 2\frac{ 3(r^2+a^2)^2
-3r_0^2r^2-4r_0^2a^2}{(r^2+a^2)^2(r^2+a^2-r_0^2))}
 \ee
is also constant for $r=$cons and it is
positive for:
\be
r > {1 \over 6}
\sqrt{18r_0^2-36a^2+6\sqrt{9r_0^4+12r_0^2a^2)}} \, .
\ee

\subsection{Flat space-time limit}

When $r_0=0$  the metric takes the form \be\n{2.2c} ds^2=- dt^2
+(r^2+a^2)\, \sin^2\theta\, d\phi^2+ (r^2+b^2)\, \cos^2\theta\,
d\psi^2 +{r^2\rho^2 \,dr^2\over (r^2+a^2)(r^2+b^2)}+\rho^2\,
d\theta^2\, . \ee
Such a space-time is flat and the metric can be
rewritten as

\be\n{2.3} ds^2=-dT^2+dX^2+dY^2+dZ^2+dW^2=-dT^2+dR^2+R^2\,
d\Omega_3^2\, , \ee where $T=t$ and
\[
X=\sqrt{r^2+a^2}\, \sin\theta\, \cos\tilde{\phi}\, ,\hspace{1cm}
Y=\sqrt{r^2+a^2}\, \sin\theta\, \sin\tilde{\phi}\, ,
\]
\be\n{2.4} Z=\sqrt{r^2+b^2}\,\cos\theta\,  \cos \tilde{\psi}\,
,\hspace{1cm} W=\sqrt{r^2+b^2}\, \cos\theta\, \sin \tilde{\psi}\,
. \ee Here \be\n{2.a} \tilde{\phi}=\phi-\tan^{-1}(a/r)\, ,
\hspace{1cm} \tilde{\psi}=\psi-\tan^{-1}(b/r)\, . \ee \be\n{2.4a}
R^2\equiv X^2+Y^2+Z^2+W^2=\rho^2\, , \ee and $d\Omega_3^2$ is the
line element on a unit 3-sphere $S^3$. At far distances $R$ and
$r$ differs only by terms of the order of $a^2/r$ and $b^2/r$.
Also normals to $R=$const and $r=$const coincide with the same
accuracy.

A two-dimensional plane $Z=W=0$ ($X-Y$-plane $\Pi_{XY}$) is a
plane of rotation, while $\Pi_{ZW}$ where $X=Y=0$ is a two-plane
orthogonal to $\Pi_{XY}$.

\section{Scalar field equation. Separation of variables}\label{s2}
\setcounter{equation}0

Let us consider a scalar massless field $\varphi$ with the action
\be W[\varphi]= -{1\over 2}\, \int\, d^5x\, \sqrt{-g}\, \left(
(\nabla\varphi)^2 +\xi R \varphi^2 \right)\, . \ee It obeys  the
following equation \be\n{2.5} \Box \varphi -\xi R\varphi= {1\over
\sqrt{-g}}\, \partial_{\mu}\left( \sqrt{-g}\, g^{\mu\nu}
\partial_{\nu} \varphi\right)-\xi R\varphi=0\, . \ee The
space-time (\ref{1.10}) is Ricci flat so that $R=0$. Let us denote
${\Box}^{*}=\rho^2\, \Box$. Then the field equation \be\n{3.3a}
{\Box}^{*}\varphi=0 \ee can be identically written in the form
\be\n{2.6} H^{AB}\,  \varphi_{,AB}+{1\over r} \,
\partial_r\left[ {\Delta\over r}
\partial_r \, \varphi \right]\,
+{1\over \sin\theta\cos\theta}\,
\partial_{\theta}\left[\sin\theta\cos\theta\,  \partial_{\theta}\, \varphi
\right]=0\, , \ee where $H^{AB}=\rho^2\, g^{AB}$. We used
GRTensor program to calculate $g^{\mu\nu}$ and $\sqrt{-g}$. The
components of $H^{AB}$ are \be\n{2.7}
H^{tt}=(a^2-b^2)\,\sin^2\theta -
{(r^2+a^2)[\Delta+r_0^2(r^2+b^2)]\over \Delta}\, , \ee \be\n{2.7a}
H^{\phi\phi}={1\over
\sin^2\theta}-{(a^2-b^2)(r^2+b^2)+b^2r_0^2\over \Delta}\, , \ee
\be H^{\psi\psi}={1\over
\cos^2\theta}+{(a^2-b^2)(r^2+a^2)-a^2r_0^2\over \Delta}\, , \ee
\be H^{t\phi}={a\, r_0^2\, (r^2+b^2)\over \Delta}\, ,
\hspace{0.5cm} H^{t\psi}={b\, r_0^2\, (r^2+a^2)\over \Delta}\, ,
\hspace{0.5cm} H^{\phi\psi}=-{ab\, r_0^2 \over \Delta}\, . \ee

The equation (\ref{2.6}) allows the separation of variables. Namely
its solution can be decomposed into modes of the form
\be\n{2.9}
\varphi \sim e^{-i\omega\,t}\, e^{im\phi}\, e^{ik\psi}\, R(r)\,
\Theta(\theta)\, .
\ee

The angular function $\Theta$ obeys the equation
\be\n{2.10}
{d\over d\theta}\left(\sin\theta\cos\theta {d\Theta\over d\theta} \right)+\left[ \lambda
-\omega^2(a^2 \sin^2\theta+b^2 \cos^2 \theta)-{m^2\over
\sin^2\theta}-{k^2\over \cos^2\theta} \right]\,
\sin\theta\cos\theta\, \Theta=0 \, .
\ee
The radial equation reads
\be\n{2.13}
{\Delta\over r}\, {d\over dr}\left[{\Delta\over r} {dR\over dr} \right] +W\, R=0\, ,
\ee
\begin{eqnarray}
W=&&\Delta \left( -\lambda+\omega^2 (r^2+a^2+b^2) +
\frac{m^2\,(a^2-b^2)}{r^2+a^2} + \frac{k^2\, (b^2-a^2)}{r^2+b^2}
\right)
 + \\ &&r_0^2\, (r^2+a^2) (r^2+b^2) \left(w-\frac{ma}{r^2+a^2} -
 \frac{kb}{r^2+b^2} \right)^2 \, .
\nonumber \end{eqnarray}

We used the freedom $\lambda\to\lambda+$const in the choice of the
separation constant $\lambda$ in order to get the angular and radial
equations in the most symmetric way. For this choice these equations
are invariant under the transformation
\be\n{2.13a}
a \leftrightarrow  b\, ,\hspace{0.5cm}
\theta \leftrightarrow  {\pi\over 2}-\theta\, ,\hspace{0.5cm}
m  \leftrightarrow k
\, .
\ee

It should be emphasized that the separability of the field
equation (\ref{3.3a}) is directly connected with the existence of the
Killing tensor for the metric (\ref{1.10})--(\ref{1.11}). The explicit
form of this Killing tensor and discussion of its properties can be
found in \cite{FrSt}.

\section{Hyperspheroidal functions}
\setcounter{equation}0

\subsection{Generic case}

The angular equation can be  rewritten for
$S(\theta)=\sqrt{\cos\theta}\, \Theta(\theta)$ as follows
\be\n{2.11} {d^2 S\over d\theta^2}+{\cos\theta\over \sin\theta}\,
{dS\over d\theta} + U_{\lambda} S=0\, , \ee \be\n{2.12}
U_{\lambda}={\tilde \lambda} -\alpha\sin^2\theta-{m^2\over
\sin^2\theta}-{k^2-1/4\over \cos^2\theta}\, , \ee where
$\alpha=(a^2-b^2)\omega^2$ and ${\tilde \lambda}=\lambda -\omega^2
b^2 +{3\over 4}$.

We discuss now properties of solutions of this equation. We are
looking for solutions which are regular at singular points of the
equation, $\theta=0$ and $\theta=\pi/2$. This condition singles
out special discrete values of $\lambda$ which we enumerate by an
integer number $\ell$. Finding eigenvalues
$\lambda_{\ell}(m,k|\alpha)$ and eigenfunctions
$S_{\ell}^{m,k}(\theta|\alpha)$ is a well defined problem.
Standard arguments show that the eigenfunctions with different
$\lambda_{\ell}(m,k|\alpha)$ are orthogonal one to another with
the proper chosen measure. Thus we write
\be\n{2.13b}
\int_0^{\pi/2}\,d\theta\, \sin\theta\,
S_{\ell}^{m,k}(\theta|\alpha)\,
S_{\ell'}^{m,k}(\theta|\alpha)=\delta_{\ell,\ell'}\, . \ee

In what follows we shall use the following normalized set of
functions

\be\n{2.14} {\cal Y}_{\ell m
k}(\theta,\phi,\psi|\alpha)={e^{im\phi+ik\psi}\over 2\pi}\,
{S_{\ell}^{m,k}(\theta|\alpha)\over \sqrt{\cos\theta}}\, . \ee We
shall also  use  the compact notation for the index, $A=\{\ell,
m,k\}$. This set of functions possess the following normalization
conditions

\be\n{2.15} \int d\gamma\, {\cal Y}_{A}\, \bar{{\cal Y}}_{A'}\,
=\delta_{AA'}\, , \ee where \be\n{2.16} \int\, d\gamma
(\ldots)=\int_0^{\pi/2}\, \sin\theta\, \cos\theta\, d\theta\,
\int_0^{2\pi}\, d\phi\, \int_0^{2\pi}\, d\psi\, (\ldots)\, ,
\hspace{0.5cm}\delta_{AA'}=\delta_{\ell \ell'}\, \delta_{mm'}\,
\delta_{kk'}\, . \ee

With $x=\cos(\theta)$, the equation (\ref{2.11}) can be rewritten
as follows \be\n{2.17} {d\over dx}\left[ (1-x^2)\, {dS\over
dx}\right] +U_{\lambda}\, S=0\, , \ee \be\n{2.18}
U_{\lambda}=\tilde{\lambda} +\alpha\, x^2 -{m^2\over 1-x^2}\,
-{k^2-1/4\over x^2}\, . \ee
This equation besides $\infty$ has three singular
points $x=0, \pm 1$ at which solutions have the following
asymptotic behavior \be\n{2.19} S(x)\sim x^{\pm k}\, ,\mbox{   at
$x=0$}\, ,\hspace{1cm} S(x)\sim (1\mp x)^{\pm m/2}\, ,\mbox{   at
$x=\pm 1$}\, . \ee For a regular solution we have \be\n{2.20}
S(x)= x^{|k|}\, (1-x^2)^{|m|/2}\, F(x)\, , \ee where $F$ is a
solution of the equation
\[
x^{2}(1- x^{2})\, F''+ ( - 2\,x^{3}\,|m| - 2\,x^{3}\,|k| +
2\,x\,|k| - 2\,x^{3})\, F'
\]
\be\n{2.21}
 \\
\left(a^2\omega^2\,x^{4} +[\lambda -
(|m|+|k|)^{2}-(|m|+|k|)]\,x^{2} + {\displaystyle \frac {1}{4}}  -
|k|\right)\, F=0 \, . \ee The regularity condition means that we
are looking for a solutions $F(x)$ which are finite at $x=0$ and
at $x=1$. This gives us the two-point boundary value problem.

\subsection{Degenerate case}

Consider now the degenerate case when   $a=b$. A solution $S$ of
(\ref{2.11}) is of the form
\be\n{3.1}
S_{\ell}^{|m|,|k|}(\theta)=\sin^{|m|}\theta \, \sqrt{\cos\theta}\,
\left[ C_1\, \cos^{|k|}\theta\, s_{\ell}^{|m|,|k|}(|k|,\theta) +
C_2\, \cos^{-|k|}\theta\, s_{\ell}^{|m|,|k|}(-|k|,\theta)
\right]\, ,
\ee
\be\n{3.2}
s_{\ell}^{|m|,|k|}(|k|,\theta)=
F(\alpha_+, \alpha_-;1+|k|; \cos^2\theta)\, .
\ee
\be
\alpha_{\pm}={|m|+1+|k|\over 2}\pm {1\over 2}\sqrt{1+\lambda}
\ee
Here
$F(a,b;c;z)$ is a hypergeometric function. The coefficients of
this function in our case posses the property
$a+b-c=|m|=$(positive) integer number. The regularity of $S$ at
$\theta=\pi/2$ requires $C_2=0$, while its regularity at
$\theta=0$ implies
\be\n{3.3}
{|m|+1+|k|\over 2}- {1\over
2}\sqrt{1+\lambda} =-\ell\, ,
\ee
where $\ell$ is an integer number.
Thus we have
\be\n{3.4}
\lambda_{\ell}(|m|,|k|)=(2\ell+|m|+|k|+1)^2-1\, .
\ee
The
corresponding eigenfunctions are the Jacobi polynomials
$P_{\ell}^{|m|,|k|}(1-2\cos^2\theta)$.
Thus
\be\n{3.5a} {\cal Y}_{\ell m
k}(\theta,\phi,\psi|\alpha=0)={e^{im\phi+ik\psi}\over 2\pi}\,
(1-z)^{|m|/2} (1+z)^{|k|/2} B_\ell^{|m|,|k|} P_\ell^{|m|,|k|} (z)
\, . \ee where $z=\cos(2\theta)=\cos \vartheta$ and
\be \label{B}
B_\ell^{|m|,|k|}=2^{-\frac{|m|+|k|-1}{2}}
\sqrt{2\ell+|m|+|k|+1}
\sqrt{\frac{\Gamma(\ell+1)\Gamma(\ell+|m|+|k|+1)}{\Gamma(\ell+|m|+1)\Gamma(\ell+|k|+1)}}\,
.
\ee

Note  that for $a=b$ the hyperspheroidal harmonics (\ref{2.14}) are
eigen functions of the operators $J_{\Phi}$, $J_{\Psi}$, and
$J^2$ determined by (\ref{ang}). Using this fact one can demonstrate
that the set (\ref{2.14}) of hyperspheroidal harmonics is complete
(see e.g. \cite{VMK,BR}).  For this set we have
$m=0,\pm 1, \pm 2 \ldots$,   $k=0,\pm 1, \pm 2 \ldots$,        $l=0,1,  2 \ldots$.

\bigskip

\section{Radial equation}
\setcounter{equation}0

It is convenient to rewrite the radial equation (\ref{2.13}) as
follows. Let's define
\be\n{4.1}
R=\left[{r\over {(r^2+a^2)(r^2+b^2)}}\right]^{1/2}\, {\cal R}\, .
\ee
Then, one has \be\n{4.2}
{d^2{\cal R}\over dr_*^2}+V\,{\cal R}=0\, , \ee where $r_*$ is a
tortoise coordinate with the property \be\n{4.5} \frac{dr_*}{dr}=
{(r^2+a^2)(r^2+b^2) \over \Delta } \, . \ee The effective
potential $V$ is\footnote{The more general setup adopted for the string theory
framework was studied in \cite{Cvetic1,Cvetic2}}:
\[
V={r^2 \Delta \left[ -\lambda
+\omega^2\, (r^2+a^2+b^2)  + \frac{m^2\, (a^2-b^2)}{r^2+a^2}
+\frac{k^2\, (b^2-a^2) }{r^2+b^2}\right] \over
(r^2+a^2)^2(r^2+b^2)^2  }+{ Z \Delta^2  \over 4 r^2(r^2+a^2)^4
(r^2+b^2)^4}
\]
\be\n{4.3}
 +{ r_0^2 r^2 \left( \omega-\frac{ma}{r^2+a^2} - \frac{k b}{r^2+b^2}
\right)^2 \over (r^2+a^2)(r^2+b^2)  }
\ee
where
$$
Z= 21r^8+ 14 r^6 (a^2+b^2)+5r^4 (a^2-b^2)^2
-10r^2a^2b^2(a^2+b^2)-3a^4b^4\, .
$$
The radial equation is evidently invariant under the transformation
\be
a\leftrightarrow b\, \hspace{1cm}
m\leftrightarrow k\, .
\ee
At the horizon, i.e. $r=r_+$, the effective potential $V$ takes
the form
\be\n{4.6} V_{hor}=(\omega
-m\Omega_a -k \Omega_b)^2\, , \ee
with $\Omega_a$ and $\Omega_b$
defined by the eq. (\ref{2.2b}). The asymptotic value of the
potential $V$ at infinity is \be\n{4.7} V_{inf}=\omega^2\, . \ee

We define two sets of solutions $\ {\cal R}_{A}^{\rm in}(r|\omega)$
and $\ {\cal R}_{A}^{\rm up}(r|\omega)$  (with $ A=\{ \ell, m, k \} $ ) by the boundary conditions
\begin{equation}  \n{4.8}
{\cal R}_{A}^{\rm in}(r|\omega)
\sim\,\cases{ \ t_{A}^{\rm in}(\omega)
 \,e^{-i\varpi  r_*}  & as $r\to r_+$\,,  \cr
\mathstrut &\cr e^{-i\omega  r_*} +
\ r_{A}^{\rm in}(\omega)
\,e^{+i \omega  r_*} & as $ r\to\infty $\,, \cr } \quad
\end{equation}
and
 \begin{equation} \n{4.9}
{\cal R}_{A}^{\rm up}(r|\omega) \sim
\, \cases{\ r_{A}^{\rm
 up}(\omega)
 \,e^{-i\varpi  r_*} +
 e^{+i \varpi  r_*}
& as $r\to r_+ $\,, \cr
\mathstrut &\cr t_{A}^{\rm up}(\omega)\,e^{+i \omega  r_*}
 &
as $r\to\infty$\,. \cr}\
 \!
\end{equation}
Here \be\n{4.10} \varpi =\omega-m\Omega_a -k \Omega_b\, . \ee
Since the eigenvalues $\lambda_{\ell}(mk|\alpha)$ are real the
functions complex conjugated  to ${\cal R}_{J}^{\rm in}$ and
${\cal R}_{J}^{\rm up}$ are also solutions. Using the constancy
of the Wronskian for various combinations of solutions of the
radial equation one gets \be\n{4.11} 1-|r_{A}^{\rm
in}(\omega)|^2={\varpi\over \omega}\, |t_{A}^{\rm in}(\omega)|^2
\, , \hspace{0.5cm} 1-|r_{A}^{\rm
up}(\omega)|^2={\omega\over\varpi}\, |t_{A}^{\rm up}(\omega)|^2
\, , \ee \be\n{4.12} \omega\,\bar{t}_{A}^{\rm up}(\omega)\,
r_{J}^{\rm in}(\omega) \,= -\varpi\,\bar{r}_{A}^{\rm
up}(\omega)\, t_{A}^{\rm in}(\omega) \, ,\hspace{0.5cm}
\omega\,t_{A}^{\rm up}(\omega) =\varpi\, t_{A}^{\rm in}(\omega)
\, . \ee

It should be emphasized that, as it was the case in $3+1$
dimensions, for certain values of $\omega$ it is possible to have
the reflection coefficients greater than one. This implies the
existence of the  so-called {\em superradiance effect}. The condition
for superradiance is
\be \label{src}  0<\omega< m\Omega_a+k\Omega_b           \, .
\ee
In this case,
it follows from (\ref{4.11}) that the reflection coefficients
$r^{{\rm in,up}}_A(\omega)$ become greater than one. This is the
consequence of the fact that the scalar field modes obeying (\ref{src})
are amplified  by the rotating black hole.

\bigskip

\section{Energy and angular momentum fluxes}
\setcounter{equation}0

Quantization of  massless fields in the 4-dimensional space-time
of a rotating black hole was discussed in
\cite{CaChHo:81,FrTh:89,MaDaOt:93, OtWi:00}  (see also
\cite{FrNo:98}). Main aspects of the quantization including the
choice of the state remain practically the same in the
5-dimensional case. For this reason we shall not repeat here all
the formal elements of the standard procedure but simply
introduce the required notations and present the final results.
We shall follow the paper \cite{OtWi:00} where necessary details
can be found.

In order to quantize the field one uses the complete set of
orthonormal complex solutions of the field equation (\ref{2.5}).
The scalar product for these solutions is defined as follows
\be\n{5.0}
< \varphi_1,\varphi_2>={i\over 2}\, \int_{\Sigma}\, \sqrt{-g}\,
(\bar{u}_{2,\mu}\, u_1-\bar{u}_{1,\mu}\, u_2)\, d\Sigma^{\mu}\, ,
\ee
where $\Sigma$ is any complete Cauchy surface. Since the scalar
product (\ref{5.0}) does not depend on the choice of $\Sigma$, it is
convenient in our case to choose $\Sigma={\cal J}^- \cup H^-$, where
${\cal J}^-$ is the past null infinity and $H^-$ is the past horizon.
(For details, see e.g. \cite{FrNo:98}).

Using solutions of the radial and angular equations we define the
following normalized solutions of the field equation (\ref{2.5})
\  \ (in coordinates $(t,r,\theta, \phi,\psi)$)

\begin{eqnarray}\n{5.1}
&& u_{\Lambda}^{\rm in}(x)=\left[{r\over {4\pi\omega \,
(r^2+a^2)(r^2+b^2)}}\right]^{1/2}\, e^{-i\omega t} {\cal R}_{\ell
m k}^{\rm in}(r|\omega) {\cal Y}_{\ell m
k}(\theta,\phi,\psi|a\omega)\, ,      \\
\n{5.2} && u_{\Lambda}^{\rm
up}(x)=\left[{r\over {4\pi |\varpi| \,
(r^2+a^2)(r^2+b^2)}}\right]^{1/2}\, e^{-i\omega t} {\cal R}_{\ell
m k}^{\rm up}(r|\omega) {\cal Y}_{\ell m
k}(\theta,\phi,\psi|a\omega)\, . \end{eqnarray}
Here ${\Lambda}=\{\omega\ell
m k \}$. The solutions $u_{\Lambda}^{\rm in}$ are defined for
$\omega >0$. The solutions $u_{\Lambda}^{\rm up}$ are naturally
defined for $\varpi >0$. For superradiant modes, when condition (\ref{src}) is satisfied,
one uses solutions $u_{-\omega \ell -m -k}^{\rm
up}$.

The unrenormalized expectation value of the stress-energy tensor
in the state of the Unruh vacuum is (see \cite{OtWi:00})
\be\n{5.4} \la U|\hat{T}_{\mu\nu}|U\ra = \sum_J\left[
\int_0^{\infty}\, d\varpi\, \coth\left({\pi\varpi\over
\kappa}\right)\, T_{\mu\nu}[u^{\rm up}_{\Lambda},\bar{u}^{\rm
up}_{\Lambda}] + \int_0^{\infty}\, d\omega\,  T_{\mu\nu}[u^{\rm
in}_{\Lambda},\bar{u}^{\rm in}_{\Lambda}] \right]\, , \ee where
$\kappa$ is the surface gravity and \be\n{5.5}
T_{\mu\nu}[u,\bar{u}]= ({1\over 2}-\xi) (u_{;\mu}\, \bar{u}_{;\nu}
+ u_{;\nu}\, \bar{u}_{;\mu}) -\xi  (u_{;\mu\nu}\, \bar{u} + u\,
\bar{u}_{;\mu\nu}) +(-{1\over 2}+2\xi)\,  g_{\mu\nu}\,
u_{;\lambda}\, \bar{u}^{;\lambda} \, . \ee Here $\xi$ is the
parameter of non-minimal coupling. For the conformal field in
5-dimensional spacetime
$\xi=3/16$ and $T_{\mu}^{\nu}[u,\bar{u}]=0$.

Let us define the following expressions for the energy and
angular momentum density fluxes at infinity, \be\n{5.6}
\varepsilon = -\la U|\hat{T}_{\mu\nu}|U\ra \xi_{t}^{\mu}\,
n^{\nu}\, , \ee \be\n{5.7} j = -\la U|\hat{T}_{\mu\nu}|U\ra
\xi_{\phi}^{\mu}\, n^{\nu}\, . \ee Here $n^{\mu}$ is a unit
vector in $r$ direction at infinity and we denoted by $
\xi_{t}^{\mu}$ and $ \xi_{\phi}^{\mu}$ the Killing vectors which
generate translation in time $t$ and rotation in an $\phi$
direction.  Substituting the asymptotics of functions
$u_{\Lambda}^{\rm in}$ and $u_{\Lambda}^{\rm up}$ into
(\ref{5.4}) one gets for the renormalized value of fluxes (see
\cite{OtWi:00}) \be\n{5.8} \varepsilon(\theta)\sim {1\over
8\pi^3\, r^3}\, \sum_{\ell, m, k}\, \int_0^{\infty}\, {\omega^2\,
d\omega\over \varpi\, {\displaystyle
(e^{2\pi\varpi/\kappa}-1)}}\, |t_{\ell m k}^{\rm up}(\omega)|^2 \,
\, {|S_{\ell}^{m,k}(\theta|\alpha)|^2 \over \cos\theta} \, , \ee
\be\n{5.9} j(\theta)\sim {1\over 8\pi^3\, r^3}\, \sum_{\ell, m,
k}\, \int_0^{\infty}\, {m\omega\, d\omega\over \varpi\,
{\displaystyle (e^{2\pi\varpi/\kappa}-1)}}\, |t_{\ell m k}^{\rm
up}(\omega)|^2 { |S_{\ell}^{m,k}(\theta|\alpha)|^2 \over
\cos\theta} \, . \ee

As we already mentioned,in the case of $a=b \neq 0$ the  angular
harmonics $S_{\ell}^{m,k}$ are the same as the angular harmonics in
the absence of rotation. However, the radial equations (and therefore
the gray-body factors) are different for these two cases.  Because of
that, the energy and angular momentum fluxes differ from those
derived in the non-rotating case.

In the degenerate case $a=b\neq 0$, we have:

\be\n{5.8a} \varepsilon(\theta)\sim {1\over 8\pi^3\, r^3}\,
\sum_{\ell, m, k}\, \int_0^{\infty}\, {\omega^2\, d\omega\over
\varpi\, {\displaystyle (e^{2\pi\varpi/\kappa}-1)}}\, |t_{\ell m
k}^{\rm up}(\omega)|^2 \,  (1-z)^{|m|}
(1+z)^{|k|} \left(B_\ell^{|m|,|k|}\right)^2 \left(P_\ell^{|m|,|k|}
(z) \right)^2 \, , \ee
\be\n{5.9a} j(\theta)\sim {1\over 8\pi^3\, r^3}\, \sum_{\ell, m,
k}\, \int_0^{\infty}\, {m\omega\, d\omega\over \varpi\,
{\displaystyle (e^{2\pi\varpi/\kappa}-1)}}\, |t_{\ell m k}^{\rm
up}(\omega)|^2 \, (1-z)^{|m|} (1+z)^{|k|}
\left(B_\ell^{|m|,|k|}\right)^2 \left(P_\ell^{|m|,|k|} (z)
\right)^2 \, . \ee where $B_\ell^{|m|,|k|}$ is defined in
(\ref{B}) and $z=\cos(2\theta)$.

By integrating over the angle variables (over 3-sphere boundary
at infinity) we obtain the expressions for the total energy and
angular moment emission

\be\n{5.10} \dot{{\bf E}}={1\over 2\pi}\, \sum_{\ell, m, k}\,
\int_0^{\infty}\, {\omega^2\, d\omega   \over \varpi\,
{\displaystyle (e^{2\pi\varpi/\kappa}-1)}}\, |t_{\ell m k}^{\rm
up}(\omega)|^2\,  , \ee \be\n{5.11} \dot{{\bf J}}={1\over 2\pi}\,
\sum_{\ell, m, k}\, \int_0^{\infty}\, {m\omega\, d\omega \over
\varpi\, {\displaystyle (e^{2\pi\varpi/\kappa}-1)}}\, |t_{\ell m
k}^{\rm up}(\omega)|^2 \, . \ee

\section{Discussion}

The results obtained in this paper are important from several
points of view. First, they allow one to better understand the
dynamics of massless scalar fields propagating near a rotating
higher dimensional black  holes. The separation of variables
which occurs in $5$-dimensional scalar field wave equation
indicates that separation which occurred in $4$-dimensional case
was not merely an accident, but it is rather a property which
follows from the symmetries of a rotating black hole.

The results are readily applicable to the brane world models
where phenomenologically valid rotating higher dimensional black
holes can exist. Non-trivial structure of rotational group in
four spatial dimensions (like the existence of two parameters of
rotation, the notion of invariant planes of rotations rather than
axis of rotation etc.) gives rise to rich phenomenology
concerning radiation from such black holes. For example, one of
the parameters of rotation could be zero, in which case the black
hole would be spinning only in one plane. This case is of
particular interest in brane world models where in the first
approximation the black holes produced in collisions of standard
model particles can spin only in the planes defined by the brane
where all the standard model fields are confined. However, if we
take a back-reaction into account, after such a black hole emits
higher dimensional graviton into the bulk, it gains the general
angular momentum which can not be described with a single
parameter of rotation. Due to stochastic nature of the black hole
evaporation, it is not likely that the black hole rotation will
spin itself up significantly in any general direction. However, this makes
the general form of rotation important to study, at least in principle.
It is interesting to mention that if both
parameters of rotation are non-vanishing but equal, the angular
equation coincides with that of a non-rotating case. For each of
these particular cases, using expressions given in eqs.
(\ref{5.8a}) and (\ref{5.9a}), we can calculate the angular
distributions as well as the total amount of energy and angular
momentum emitted by a black hole.

We demonstrated that similarly to the 4-dimensional case,
5-dimensional rotating black hole allows the {\em superradiance
effect}. If $m$ and $k$ are azimuthal quantum numbers with respect to
2 rotation axes, then the condition of superradiance is $\omega<
m\Omega_a+k\Omega_b$, where $\Omega_a$ and $\Omega_b$ are angular
velocities. One can expect that another feature of quantum radiation
from 4-dimensional rotating black holes, namely its strong spin
dependence \cite{Page:76}, is also present in the 5-dimensional case.
If it occurs then the bulk radiation of gravitons might be the
leading channel of the black hole decay. In order to check this
conjecture it is necessary to solve higher spin massless field
equations in the higher dimensional space-time of a rotating black
hole. This is an interesting challenge.

\bigskip

\vspace{12pt} {\bf Acknowledgments}:\ \  The authors are greatful
to Don Page for stimulating discussions. This work was partly
supported  by  the Natural Sciences and Engineering Research
Council of Canada. The authors are grateful to the Killam Trust
for its financial support.

\bigskip




\begin{thebibliography}{9}




\bibitem{Tan} F. R. Thangherlini, {\it Nuovo Cimento} {\bf 77} 636
(1963)




\bibitem{BW}
N. Arkani-Hamed, S. Dimopoulos and G. Dvali, Phys. Lett. {\bf
B429}, 263 (1998); I. Antoniadis, N. Arkani-Hamed,  S. Dimopoulos
and G. Dvali, Phys. Lett. {\bf B436}, 257 (1998);
N. Kaloper, J. March-Russel, G. Starkman and M. Trodden, {\it %
Phys. Rev. Lett.} {\bf 85}, 928 (2000);    G. Starkman, D.
Stojkovic and M. Trodden,  Phys. Rev. Lett. {\bf 87} 231303
(2001); Phys. Rev.  {\bf D63}, 103511 (2001); L. Randall and R.
Sundrum, Phys. Rev. Lett. {\bf 83}, 3370 (1999); {\it ibid} 4690
(1999);


\bibitem{acc}
S. Dimopoulos, G. Landsberg, Phys. Rev. Lett. {\bf 87} 161602
(2001);
 D. M. Eardley, S. B. Giddings, {\it gr-qc/0201034} ;
 S. B. Giddings, S. Thomas, Phys. Rev. {\bf D65} 056010 (2002);
 J.  L. Feng, A. D. Shapere,  Phys. Rev. Lett. {\bf 88} 021303
 (2002);
 G. Landsberg  {\it hep-ph/0211043} ; hep-ph/0205174;
S. B. Giddings, hep-th/0205027 ;  V. Frolov, D. Stojkovic, Phys.
Rev. Lett. {\bf 89} 151302 (2002); Phys. Rev. {\bf D66} 084002
(2002);  M.B. Voloshin , Phys. Lett.  {\bf B524} 376 (2002 );
 S. Solodukhin, Phys. Lett. {\bf B533} 153
(2002); R. Emparan, G. Horowitz, R. C. Myers, Phys. Rev. Lett.
{\bf 85} 499 (2000); M. Bleicher, {\it hep-ph/0112186 };   A.
Ringwald, hep-ph/0212342; A. Ringwald, H. Tu,   Phys. Lett. {\bf
B525} 135 (2002); M.~Kowalski, A.~Ringwald and H.~Tu, Phys.\
Lett.\ B {\bf 529} (2002) 1 ; H.~Tu, hep-ph/0205024;
hep-ph/0211159 ; Y. Uehara , {\it hep-ph/0110382 }; {\it
hep-ph/0203244};
R. Casadio, B. Harms, Phys. Lett. {\bf B487} 209 (2000) ; Phys.
Rev. {\bf D64}   024016 (2001); Int.J.Mod.Phys. {\bf A17} 4635
(2002); D. Stojkovic hep-ph/0111061; V. Cardoso, J. P.S. Lemos,
gr-qc/0211094; Phys. Lett. {\bf B538} 1 (2002);
  L. Anchordoqui, H. Goldberg, Phys. Rev. {\bf D65} 047502
(2002);hep-ph/0209337 ; L. A. Anchordoqui, J. L. Feng, H. Goldberg
and A. D. Shapere, hep-ph/0207139; L. A.  Anchordoqui, T. Paul,
S. Reucroft
and J. Swain, hep-ph/0206072; L. A. Anchordoqui, J. L. Feng, H.
Goldberg and A. D. Shapere, Phys. Rev. {\bf D65} 124027 (2002);
T. G. Rizzo,  JHEP  0202:011 (2002);   P. Kanti, J.March-Russell.
 hep-ph/0212199 ; Phys. Rev. {\bf D66} 024023 (2002);
 K. Cheung, Phys. Rev. Lett. {\bf 88} 221602 (2002);
hep-ph/0205033; A. Chamblin and G. C. Nayak, hep-ph/0206060;
G. C. Nayak, hep-ph/0211395 ; T. Han, G. D. Kribs and B. McElrath,
hep-ph/0207003;
 R.Emparan, R.W.Gregory and C.Santos, Phys. Rev. {\bf D63}
104022 (2001);
 M.Rogatko, Phys. Rev {\bf D64}  064014  (2001)
S. C. Park, K. Oda, D. Ida, hep-th/0212108 ; D. Ida, Y. Uchida,
Y. Morisawa  gr-qc/0212035.; R. Emparan , A. Fabbri, N. Kaloper
 JHEP {\bf 0208} 043  (2002); R. Emparan , J. Garcia-Bellido,
 N. Kaloper, hep-th/0212132;




\bibitem{GibHer}  G. W. Gibbons and C. A. R. Herdeiro,
Class. Quant. Grav. {\bf 16} 3619 (1999).


\bibitem{Her}     C. A. R. Herdeiro,
Nucl. Phys. {\bf B582} 363 (2000).


\bibitem{Cvetic1}  M. Cvetic and F. Larsen,
Phys. Rev. {\bf D56} 4994 (1997).

\bibitem{Cvetic2}  M. Cvetic and  D. Youm.
Nucl. Phys. {\bf B476}  118 (1996)


\bibitem{MyPe:86} R. C. Myers and M. J. Perry, {it Ann. of Phys.}
{\bf 172} 304 (1986)


\bibitem{Galtsov} G. Clement,  D. Gal'tsov and C. Leygnac, hep-th/0208225.

\bibitem{Petrov} A. Z. Petrov. {\em Einstein spaces}, Oxford, New
York, Pergamon Press (1969).

\bibitem{Geroch} R. Geroch, {\em Journ. Math. Phys.} {\bf 12} 918
(1971).

\bibitem{FrSt} V. Frolov, D. Stojkovic, gr-qc/0301016.

\bibitem{VMK} D. A. Varshalovich, A. N. Moskalev, and V. K.
Khersonskii. {\em Quantum Theory of Angular Momentum}, Singapore,
World Scientific (1988).

\bibitem{BR} A. O. Barut and R. Raszka. {\em Theory of Group
Representations and Applications}, Singapore,
World Scientific (1986).

\bibitem{CaChHo:81} P. Candelas, P. Chrzanowski and K. W. Howard, {\em
Phys. Rev.} {\bf D24} 297 (1981).

\bibitem{FrTh:89} V.P. Frolov and K. S. Thorne,  {\em
Phys. Rev.} {\bf D39} 2125 (1989).

\bibitem{MaDaOt:93} A. L. Matacz, P. C. W. Davies and A. C. Ottewill,  {\em
Phys. Rev.} {\bf D47} 1559 (1993).

\bibitem{OtWi:00} A. C. Ottewill and E. Winstanley, {\em Phys.Rev.}
{\bf D62} 084018  (2000).

\bibitem{FrNo:98} V. Frolov and I. Novikov. {\em Black Hole Physics: Basic
Concepts and New Developments} (Kluwer Academic Publ.), 1998.


\bibitem{Page:76} D. N. Page, Phys. Rev. {\bf D13} 198 (1976);
    Phys. Rev. {\bf D14} 3260 (1976).


\end{thebibliography}
\end{document}